\documentclass[]{spie}  

\usepackage{amsmath,amsfonts,amssymb}
\usepackage{graphicx}
\usepackage[colorlinks=true, allcolors=blue]{hyperref}
\def\arcsec{\hbox{$^{\prime\prime}\:$}}
\def\arcmin{\hbox{$^\prime\:$}}

\title{The Wide Integral Field Infrared Spectrograph: Commissioning Results and On-sky Performance}

\author[a,b]{Suresh Sivanandam}
\author[a]{Dae-Sik Moon}
\author[a]{R. Elliot Meyer}
\author[b]{Jason Grunhut}
\author[c]{Dennis Zaritsky}
\author[c]{Joshua Eisner}
\author[a]{Ke Ma}
\author[d]{Charles Henderson}
\author[e]{Basil Blank}
\author[f]{Chueh-Yi Chou}
\author[g,h]{Miranda E. Jarvis}
\author[i]{Stephen Eikenberry}
\author[j]{Moo-Young Chun}
\author[j]{Byeong-Gon Park}

\affil[a]{Department of Astronomy and Astrophysics, University of Toronto, 50 St. George St, Toronto, ON, Canada}
\affil[b]{Dunlap Institute of Astronomy and Astrophysics, University of Toronto, 50 St. George St, Toronto, ON, Canada}
\affil[c]{Steward Observatory, University of Arizona, 933 N. Cherry Ave, Tucson, AZ, USA}
\affil[d]{Department of Astronomy, Cornell University, 616A Space Science Building, Ithaca, NY, USA}
\affil[e]{PulseRay Inc., 4583 State Route 414, Beaver Dams, NY, USA}
\affil[f]{Academia Sinica - Institute of Astronomy and Astrophysics, No.1, Sec. 4, Roosevelt Rd, Taipei, Taiwan}
\affil[g]{Max Planck Institute for Astrophysics, Karl-Schwarzschild-Strasse 1, Garching, Germany}
\affil[h]{European Southern Observatory, Karl-Schwarzschild-Strasse 2, Garching, Germany}
\affil[i]{Department of Astronomy, University of Florida, 211 Bryant Space Science Center, Gainesville, FL, USA}
\affil[j]{Korea Astronomy and Space Science Institute, 776 Daedeokdae-ro, Yuseong-gu, Daejeon, Korea}

\authorinfo{Further author information: (Send correspondence to S.S.)\\S.S.: E-mail: sivanandam@dunlap.utoronto.ca, Telephone: +1 (416) 978-6550}

\begin{document} 
\maketitle

\begin{abstract}
We have recently commissioned a novel infrared ($0.9-1.7$ $\mu$m) integral field spectrograph (IFS) called the Wide Integral Field Infrared Spectrograph (WIFIS). WIFIS is a unique instrument that offers a very large field-of-view (50\arcsec x 20$^{\prime\prime}$) on the 2.3-meter Bok telescope at Kitt Peak, USA for seeing-limited observations at moderate spectral resolving power. The measured spatial sampling scale is $\sim1\times1$\arcsec and its spectral resolving power is $R\sim2,500$ and $3,000$ in the $zJ$ ($0.9-1.35$ $\mu$m) and $H_{short}$ ($1.5-1.7$ $\mu$m) modes, respectively. WIFIS's corresponding etendue is larger than existing near-infrared (NIR) IFSes, which are mostly designed to work with adaptive optics systems and therefore have very narrow fields. For this reason, this instrument is specifically suited for studying very extended objects in the near-infrared such as supernovae remnants, galactic star forming regions, and nearby galaxies, which are not easily accessible by other NIR IFSes. This enables scientific programs that were not originally possible, such as detailed surveys of a large number of nearby galaxies or a full accounting of nucleosynthetic yields of Milky Way supernova remnants. WIFIS is also designed to be easily adaptable to be used with larger telescopes. In this paper, we report on the overall performance characteristics of the instrument, which were measured during our commissioning runs in the second half of 2017. We present measurements of spectral resolving power, image quality, instrumental background, and overall efficiency and sensitivity of WIFIS and compare them with our design expectations. Finally, we present a few example observations that demonstrate WIFIS's full capability to carry out infrared imaging spectroscopy of extended objects, which is enabled by our custom data reduction pipeline.
\end{abstract}


\section{INTRODUCTION}
\label{sec:intro}  

Wide-field, seeing-limited, visible-light integral field spectroscopy has now become prevalent and it is breathing life into $2-4$-meter class telescopes by enabling large surveys of thousands of nearby ($z\leq0.1$) galaxies, which are revolutionizing the field of galaxy formation and evolution (e.g. CALIFA\cite{sanchez2012}, MaNGA\cite{bundy2015}, and SAMI\cite{croom2012}). The ability to spatially resolve spectral signatures of ionized gas and stars in two dimensions (2D) allows the detailed study of star formation, stellar populations, gas ionization states, and kinematics of a multitude of astrophysical objects. Consequently, several instruments have left a strong scientific legacy. For example, the lenslet-based integral field spectrograph (IFS), SAURON\cite{bacon2001}, on the William Herschel Telescope, enabled the SAURON and ATLAS3D surveys that made major contributions towards understanding the properties and structure of nearby galaxies. And now MUSE on the Very Large Telescope\cite{bacon2010}, which is unrivalled in its capability to spatially and spectrally resolve astronomical objects in the visible, is currently making a broad scientific impact in topics ranging from stellar evolution to galaxy formation.
\par
The ability to carry out wide-field infrared integral field spectroscopy remains limited. Existing infrared IFSes focus on adaptive optics-fed high spatial resolution and narrow-field imaging spectroscopy (e.g., SINFONI/SPIFFI\cite{eisenhauer2000}, OSIRIS\cite{larkin2006}). While this is particularly well-suited for studying central regions of the Milky Way and nearby galaxies, as well as distant high redshift galaxies, the fields-of-view are not well-matched to complement existing optical surveys of nearby galaxies or observe very extended objects such as supernova remnants within our galaxy. The near-infrared offers a powerful view of supernova remnants where emission features can help constrain nucleosynthetic yields. Moreover, the near-infrared rest-frame spectra of galaxies, which are largely unaffected by extinction, consist of important signatures of star formation and the reddest (evolved giants and low-mass dwarfs) stars. 
\par
In this paper, we present the details and performance of a recently commissioned infrared IFS that is specifically tailored to observe large fields at the seeing limit. This instrument, called the Wide Integral Field Infrared Spectrograph (WIFIS)\footnote{The principal investigator of WIFIS is Dae-Sik Moon.}, is the widest field infrared IFS that currently exists, which allows spectroscopy at moderate spectral resolving power ($R\sim3,000$). It is also the largest etendue ($A\Omega$ product) infrared IFS\cite{sivanandam2012}, allowing efficient observations of very extended objects. The original concept and design details for this instrument can be found in previous work\cite{chou2010,sivanandam2012,meyer2016}. This powerful new instrument will be able to image large sections of nearby galaxies in a single shot and complement current visible-light IFS surveys of nearby galaxies. We organize the paper as follows: in Section \ref{sec:science}, we discuss the scientific drivers for the instrument; in Section \ref{sec:instrument}, we outline the overall design and measured parameters for the instrument; in Section \ref{sec:technique}, we highlight the calibration and observation techniques we have developed for WIFIS; in Section \ref{sec:pipeline}, we summarize the features of the WIFIS data reduction pipeline; in Section \ref{sec:performance}, we present the overall performance of the instrument and example results and outstanding performance issues; and finally in Section \ref{sec:conclusions}, we summarize our findings.

\section{SCIENCE DRIVERS}
\label{sec:science}
The unique combination of the large integral field size, wavelength coverage, and spectral resolving power of WIFIS opens up entirely new scientific areas of research for investigating dynamics and chemistry of extended objects in the near-infrared (NIR) regime. WIFIS is especially well-equipped to conduct spectral mapping of an extended area (e.g., sections of the Orion Nebula or supernova remnants) or to follow-up targets observed by wide-field, visible-light, IFS surveys (e.g., CALIFA \& MaNGA). We outline two galactic and two extragalactic scientific programs that we are currently pursuing.

\subsection{Galactic: Young Supernova Remnants}
Young supernova remnants (SNRs) are rich in NIR lines, which serve as diagnostics of supernova nucleosynthetic processes and explosions. This is especially true for the $zJ$-band that has transitions of iron, sulphur, phosphorus, oxygen, carbon, and helium, which have crucial information for the supernova nucleosynthesis and explosion mechanisms. WIFIS observations will make it possible to conduct precise abundance comparisons of the supernova nucleosynthetic elements. For instance, we have recently discovered an increased abundance of phosphorus in Casseopia A\cite{koo2013}, confirming the {\it in-situ} creation of phosphorus - one of the six indispensable elements of human bodies. WIFIS observations will provide a comprehensive view of the distribution of the nucleosynthetic elements and information for their abundance and kinematics in SNRs. This will enable a thorough investigation into the nucleosynthesis and explosion dynamics of core-collapse supernovae.

\subsection{Galactic: Young Stellar Objects in Star Forming Regions}
The competitive multiplexing capability of WIFIS is well-suited to the study of clustered young stellar objects (YSOs) in star forming regions. WIFIS will carry out multi-epoch mapping of a $\sim$ $2\times2$\arcmin area of the Orion Nebular Cluster where there is a strong aggregation of YSOs. Using Pa$\beta$, [Fe II] and He I lines, the accretion and outflow rates of YSOs can be quantified. The multi-epoch observations will help constrain the variability of accretion processes from these roughly coeval sources in a young cluster. These processes relate directly to star formation since accretion rate determines the mass growth rate of the star and to planet formation via improved understanding of disk viscosity. WIFIS will also observe proplyds i.e., sources surrounded by cometary emission characteristic of offset ionization fronts. These observations will provide many diagnostic lines (e.g., Pa$\beta$, [Fe II]) to investigate the star-forming activity, especially outflow interaction with the surrounding medium, in 2D and will be complementary to the existing data in other wavebands.

\subsection{Extragalactic: Initial Mass Functions in Bulges and Ellipticals}
The goal of this program is to use NIR absorption line diagnostics to study the initial mass function (IMF) of nearby bulges of spirals and ellipticals. The NIR offers a unique window for constraining the dwarf-to-giant ratio in galactic spectra because it contains numerous stellar absorption features (e.g., NaI, CN, FeH Wing-Ford) that are sensitive to either dwarfs or giants. There has been recent visible-light work\cite{conroy2012} that suggests that the IMF becomes increasingly bottom-heavy as the velocity dispersion (i.e., mass) of the bulge increases. The most massive ellipticals exhibit an IMF that is much more bottom-heavy than Salpeter. CALIFA observations of visible-light IMF tracers also suggest a strong correlation with local metallicity\cite{martin-navarro2015}. This is quite surprising and controversial, and there is concern that the line diagnostics used in these works are affected by incomplete metallicity coverage in the spectrophotometric stellar data used in the population synthesis models.  WIFIS observations of the dwarf/giant specific absorption features, combined with visible-light diagnostics obtained from the public CALIFA and MaNGA data, can constrain the giant/dwarf ratio, and hence the low-end IMF. This, in turn, relates to the baryonic and dark matter distributions in these galaxies when kinematics are included, which would allow the determination of the root cause of IMF variations. 

\subsection{Extragalactic: Massive Star Formation and Nuclear Activity in Nearby Galaxies}
With the aid of NIR emission line diagnostics and continuum emission available from WIFIS observations, the nature of massive star formation can be studied by observing nearby star-forming galaxies. Using He I, [Fe II], Pa$\beta$, and infrared continuum in the $zJ$-band as diagnostics, WIFIS will enable the age dating of young stellar clusters in the galaxies by constraining the OB stellar fraction while also addressing the longstanding puzzle of the preferred locations for massive star formation: do they form in clusters or in the field? This will also allow the investigation of the evolution of star forming regions under the influence of the strong spiral pattern and locally generated massive star feedback in these galaxies. Comparison between the supernova rate and star forming rate, which can be traced by [Fe II] and Pa$\beta$ lines\cite{rosenberg2012}, may allow a spatial and temporal mapping of formation versus death (or feedback) of massive stars in these galaxies\cite{boker2008}. The effects of the AGN on its surroundings can also be explored by using multiple high excitation lines of silicon, sulphur and others, which will provide diagnostics of the radiation field strength and mechanical energy input via shocks in the nuclear region.

\section{INSTRUMENT DESCRIPTION}
\label{sec:instrument}
\begin{figure}[h]
\centering
\includegraphics[width=13cm]{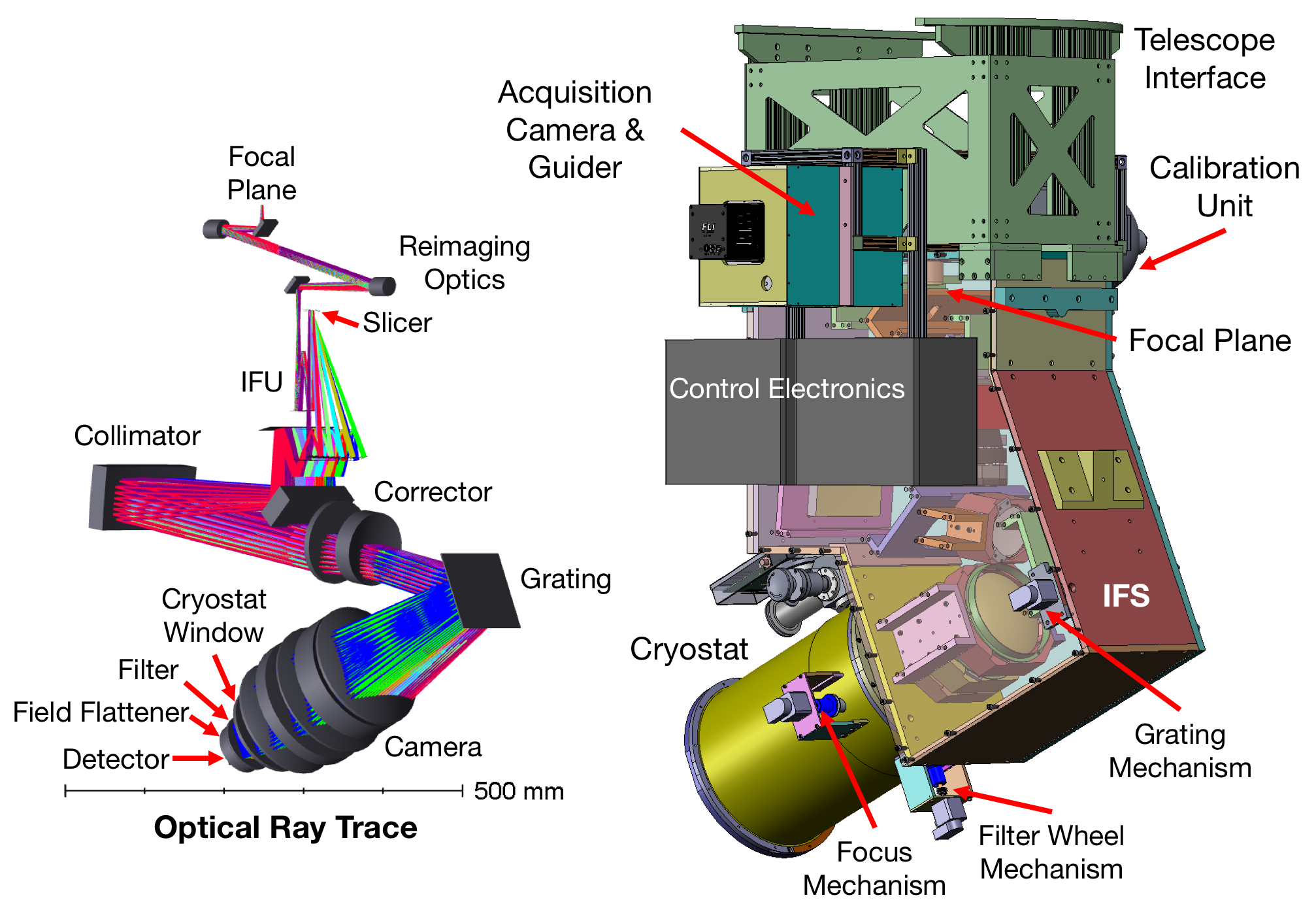}
\caption{{\it Left:} Optical layout and ray trace for WIFIS. The telescope focal plane is shown at the top of this diagram. The field is then reimaged onto an image slicer-based integral field unit (IFU) called FISICA\cite{eikenberry2006}. The reimaging optics can be replaced to accommodate different telescopes. This IFU forms a pseudoslit, which is then dispersed by the spectrograph optics downstream of it. Only the final two optical elements (filter and field flattener lens) and focal plane array (H2RG detector) are cryogenically (77K) cooled. {\it Right:} Full CAD rendering of WIFIS and its subsystems, which include an acquisition/guiding camera and a calibration unit consisting of flatfield and wavelength calibration sources. The CAD rendering is oriented to match the optical ray trace. There are three moving mechanisms in the integral field spectrograph, two of which operate cryogenically. The grating mechanism allows for a minor change in grating tilt to accommodate the two spectral modes ($zJ$ and $H_{short}$) and refine the wavelength range required in the spectrograph. The two cryogenic mechanisms are for a filter wheel, which selects between a blank, $zJ$, and $H$ filters, and a detector focusing mechanism.}
\label{fig:layout}
\end{figure}

\begin{figure}[h]
\centering
\includegraphics[width=5cm]{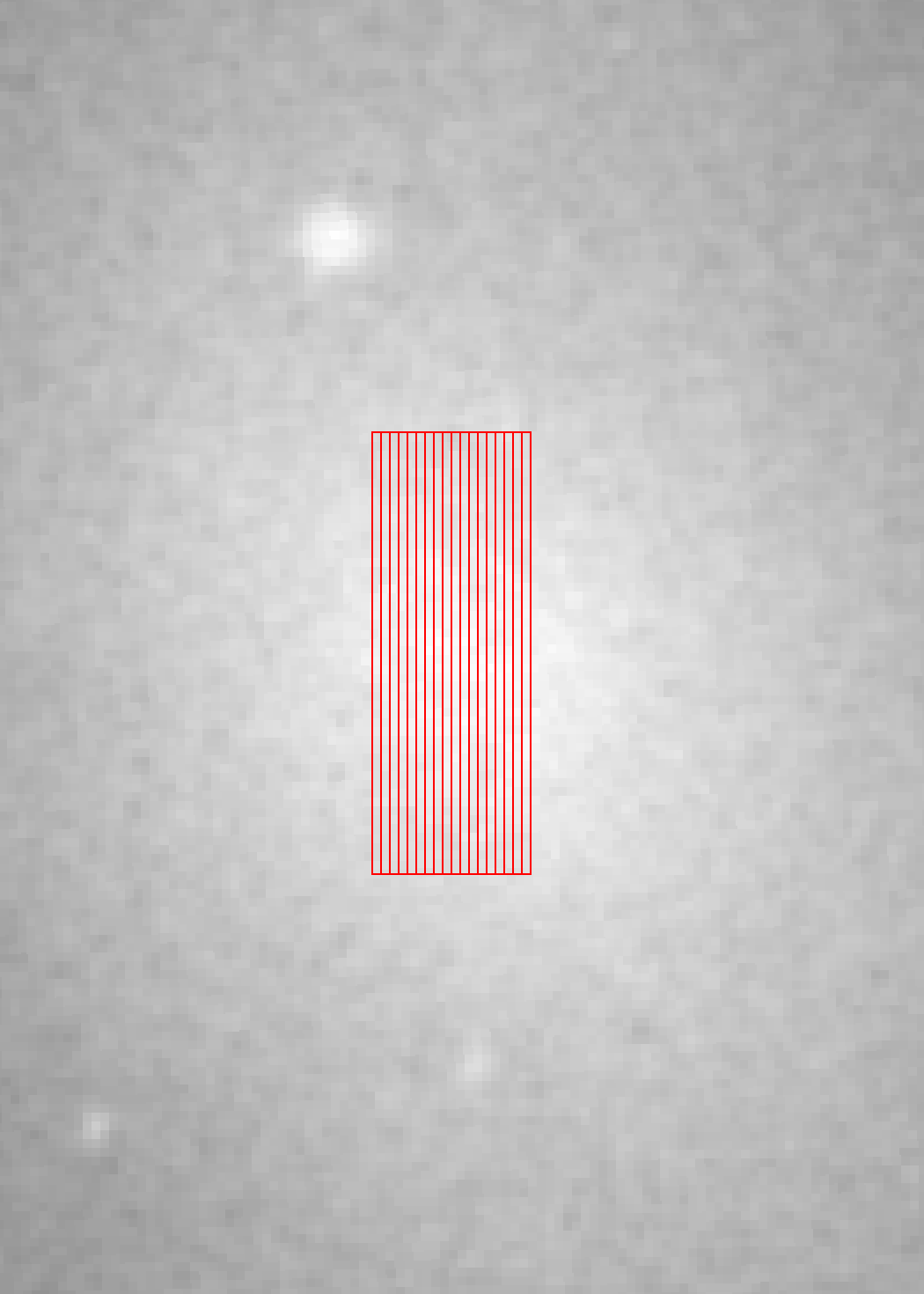} \hspace{1cm}\raisebox{-1.0cm}{\includegraphics[width=5cm]{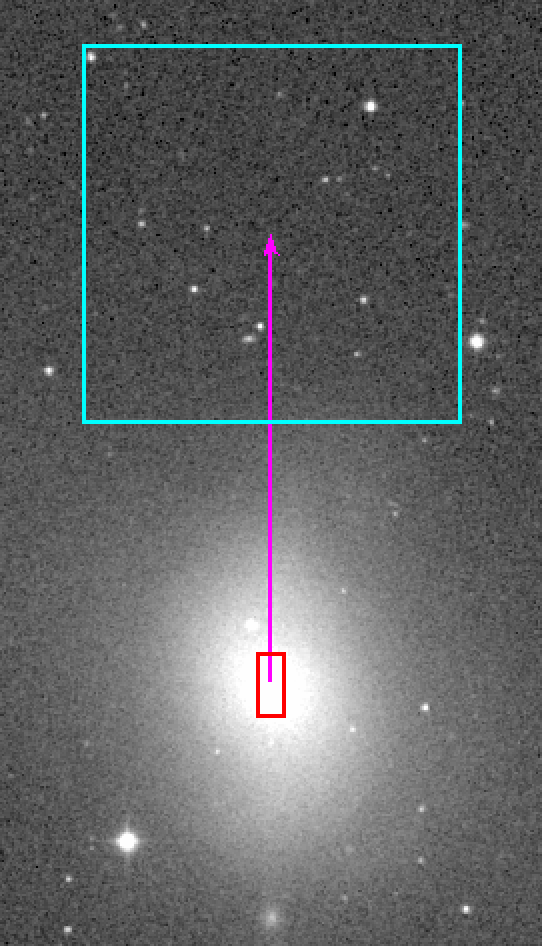}}
\caption{{\it Left:} A digitized sky survey (DSS) image of the central region of Messier 85 with the WIFIS ($50\times20^{\prime\prime}$) overlaid. The individual slices (18 total) and their orientation are shown in this image. Given that WIFIS is a slicer-based IFS, we are able to resolve spatial information along the direction of each slice. {\it Right:} The same larger field DSS image of M85 is shown with the WIFIS full field (red) and AGC (cyan) field. The AGC field is $5\times5$\arcmin in size. The centres of the two fields are almost exactly separated by 6\arcmin along the slice direction. }
\label{fig:pointing}
\end{figure}

In this work, we present the as-built details of the overall instrument. WIFIS consists of three major subsystems: the infrared integral field spectrograph (IFS), the acquisition and guider camera (AGC), and the calibration unit (CU). This instrument is mounted on the Cassegrain focus of the Steward 2.3-meter Bok telescope at Kitt Peak, USA. The critical aspects of WIFIS's optical and mechanical design details are discussed elsewhere\cite{chou2010,sivanandam2012,meyer2016}. For reference, we show the overall optical and mechanical layout of WIFIS in Figure \ref{fig:layout}. Tables \ref{tab:wifis} and \ref{tab:guider} provide the details of the integral field spectrograph and the AGC respectively. The AGC and CU play a crucial supporting role for the IFS. The AGC, which has an offset field from WIFIS, is used to acquire targets and centre them within the smaller WIFIS field. Figure \ref{fig:pointing} shows their respective fields-of-view and the 6\arcmin offset between the centres of their fields. Once the target is acquired, the AGC is then used to measure pointing corrections by guiding on bright sources. The CU consists of two selectable light sources: a thorium-argon (ThAr) cold cathode lamp used to generate light for wavelength calibration, and an integrating sphere with a quartz lamp for flatfielding. The CU has optical components to simulate the f/9 beam generated by the telescope. The CU is also used for the spatial calibration of the IFS through the insertion of a Ronchi mask. We discuss more details of how the acquisition and calibrations are done in Section \ref{sec:technique}.

\begin{table}[htb]
\caption{Integral Field Spectrograph Parameters} 
\label{tab:wifis}
\begin{center}       
\small
\begin{tabular}{|l|l|} 
\hline
\rule[-1ex]{0pt}{3.5ex}  {\bf Telescope} & Steward 2.3-meter Bok Telescope   \\
\hline
\rule[-1ex]{0pt}{3.5ex}  {\bf Location} & Kitt Peak, USA  \\
\hline
\rule[-1ex]{0pt}{3.5ex}  {\bf Field-of-View} &  $50\times20$\arcsec  \\
\hline
\rule[-1ex]{0pt}{3.5ex}  {\bf Number of Slices} &  18  \\
\hline
\rule[-1ex]{0pt}{3.5ex}  {\bf Slice Scale} &  $1.1$\arcsec  \\
\hline
\rule[-1ex]{0pt}{3.5ex}  {\bf Spaxel Scale (2 pixel sampling)} & $\sim1.1\times1.1$\arcsec   \\
\hline
\rule[-1ex]{0pt}{3.5ex}  {\bf Operating Modes} & $0.9-1.35\mu m$ ($zJ$) \& $1.5-1.7\mu m$ ($H_{short}$)  \\
\hline
\rule[-1ex]{0pt}{3.5ex}  {\bf Spectral Resolving Power (R)} & 2,500 ($zJ$) \& 3,000 ($H_{short}$)    \\
\hline
\rule[-1ex]{0pt}{3.5ex}  {\bf Dispersion (nm pix$^{-1}$)} & 0.212   \\
\hline
\rule[-1ex]{0pt}{3.5ex}  {\bf Background (e$^-$ s$^{-1}$ pix$^{-1}$)} & $2-10$ ($zJ$) \&  $100-500$ ($H_{short}$) \\
\hline
\rule[-1ex]{0pt}{3.5ex}  {\bf Detector} & Teledyne $2048\times2048$ HAWAII-2RG (H2RG) 1.7$\mu$m-cutoff   \\
\hline
\rule[-1ex]{0pt}{3.5ex}  {\bf Focal Plane Electronics} & Teledyne SIDECAR ASIC + JADE2   \\
\hline
\rule[-1ex]{0pt}{3.5ex}  {\bf Operating Temperature} & 77K \\
\hline
\rule[-1ex]{0pt}{3.5ex}  {\bf Number of Channels} & 32 (1.45479s single frame readout)  \\
\hline
\rule[-1ex]{0pt}{3.5ex}  {\bf Full Well Depth (e$^{-}$)} & 90,000   \\
\hline
\rule[-1ex]{0pt}{3.5ex}  {\bf Conversion Gain (e$^{-}$/ADU}) & 1.33   \\
\hline
\rule[-1ex]{0pt}{3.5ex}  {\bf Readout Noise (e$^{-}$)} &  30 (CDS) \& 7 (200 Up-The-Ramp reads) \\
\hline
\end{tabular}
\end{center}
\end{table} 

\begin{table}[htb]
\caption{Acquisition and Guider Camera Parameters} 
\label{tab:guider}
\begin{center}       
\small
\begin{tabular}{|l|l|} 
\hline
\rule[-1ex]{0pt}{3.5ex}  {\bf Field-of-View} &  $4.97\times4.97$\arcmin  \\
\hline
\rule[-1ex]{0pt}{3.5ex}  {\bf Plate Scale} &  0.2913\arcsec pix$^{-1}$ \\
\hline
\rule[-1ex]{0pt}{3.5ex}  {\bf Imaging Bands} & $g,r,i,z,$ and $H\alpha$  \\
\hline
\rule[-1ex]{0pt}{3.5ex}  {\bf Detector} & e2v $1024\times1024$ CCD47-20 Frame Transfer CCD    \\
\hline
\rule[-1ex]{0pt}{3.5ex}  {\bf Focal Plane Electronics} & Finger Lakes Instrumentation ML4720   \\
\hline
\rule[-1ex]{0pt}{3.5ex}  {\bf Operating Temperature} & $-30$C \\
\hline
\rule[-1ex]{0pt}{3.5ex}  {\bf Number of Outputs} & 2   \\
\hline
\rule[-1ex]{0pt}{3.5ex}  {\bf Dark Current (e$^-$ s$^{-1}$ pix$^{-1}$)} & $\sim 0.5$ \\
\hline
\rule[-1ex]{0pt}{3.5ex}  {\bf Full Well Depth (e$^{-}$)} & 96,000   \\
\hline
\rule[-1ex]{0pt}{3.5ex}  {\bf Conversion Gain (e$^{-}$/ADU}) & 1.49  \\
\hline
\rule[-1ex]{0pt}{3.5ex}  {\bf Readout Noise (e$^{-}$)} &  16 (Fast mode) \& 9 (Slow mode) \\
\hline
\end{tabular}
\end{center}
\end{table}

\begin{figure}[h]
\centering
\includegraphics[width=13cm]{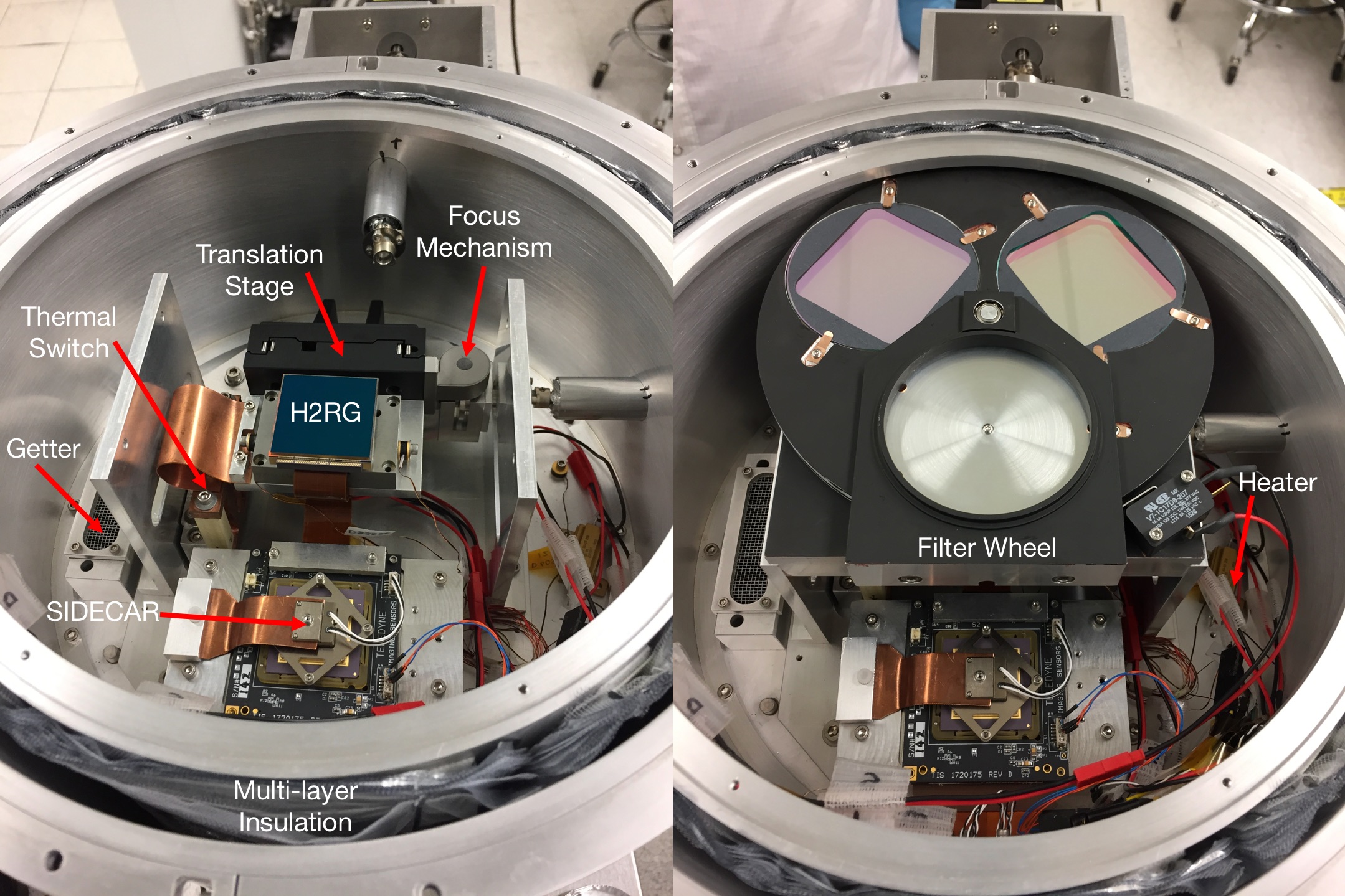}
\caption{The internal components of the WIFIS cryostat are shown here. The radiation shield has been removed to expose the components inside. The base to which all components are attached to is the cold plate, which is directly connected to a LN2 reservoir. {\it Left:} This shows the focal plane assembly of WIFIS, including the H2RG detector. The H2RG is mechanically attached to a linear translation stage that is used for focussing. This stage has temperature sensors and heaters to adjust the temperature of the detector. The H2RG is electrically connected to the SIDECAR ASIC cold electronics, which clock the array and digitize its output. Another flex cable (not shown here) connects the SIDECAR to the warm electronics outside the cryostat that communicates with the control computer. {\it Right:} The three-position filter wheel, which also includes the field flattener lens (not seen here), has been installed in the cryostat. The filter currently shown in position is the blank filter, which is used to take darks. Aeroglaze Z306 paint was used to reduce any scattered light from entering the inside the cryostat, which mainly originates from the thermal background. The 30W heater shown is used to warm up the cryostat at the end of each run.}
\label{fig:cryostat}
\end{figure}

The IFS has a unique design where most of its optics are uncooled. This significantly reduces the cost and complexity of the overall instrument. The heart of the IFS is an image slicer-based IFU called FISICA\cite{eikenberry2006}, which consists of optical elements that have been diamond-turned from 6061 aluminum. The IFU converts a rectangular field into a pseudoslit, which is then dispersed by the spectrograph optics located downstream. The final portion of the optical system is located inside a cryostat to mitigate the effect of thermal emission. Figure \ref{fig:cryostat} shows the internal structure and layout of the cryostat. The optical filter that determines the operating band of the instrument and the field flattener lens in the camera are housed in the cryostat that is cooled by liquid nitrogen (LN2) and held at 77K. Switching between the $zJ$ and $H_{short}$ modes is accomplished by shifting the grating angle by a few degrees, selecting the appropriate filter inside the cryostat, and adjusting the focus of the infrared array. The filters available are a specially designed $zJ$-band filter, manufactured by Asahi Spectra, with high suppression of the thermal background in the $1.35-3.0$ $\mu$m band, and an MKO $H$-band filter. Thermal light suppression is an especially important issue in this instrument's design as the cryostat can see thermal light over an f/1 beam. We discuss the efficacy of our thermal light suppression methods and future directions in greater detail in Section \ref{sec:performance}. 
\par
The HAWAII-2RG (H2RG) infrared array is mounted on a specially designed cryogenized translation stage that allows small focus adjustments in  2.5$\mu$m steps. During the initial cool-down of the cryostat, the detector is thermally decoupled from the cold plate to prevent it from cooling at too fast a rate ($>1$K min$^{-1}$). When the temperature reaches 170K, however, a passive thermal switch based on differential contraction (G10/copper structure) closes and provides a strong thermal connection between the focal plane assembly and the cold plate. No additional heating is required to maintain a proper cooling rate for the detector. Nevertheless, we have two high precision temperature sensors that monitor the detector temperature continuously, and there are associated heaters on the focal plane assembly that can be used if necessary. The readout of the detector is carried out by the SIDECAR ASIC cold electronics that clocks and digitizes the output of the array. A cryogenic flex cable connects the SIDECAR to the JADE2 warm electronics board outside the cryostat. A custom-built low noise power supply powers the detector system, and the system communicates over the USB bus with the WIFIS control computer. The USB connection is electrically isolated from the computer through a USB/fibre-optic coupler. We use the stock Windows-based readout software provided by Teledyne. The majority of the control software operates under a Linux computer, so the H2RG control software runs on a Windows virtual machine that communicates with and saves data on the Linux host machine.

\section{OBSERVING METHODS}
\label{sec:technique}
Observing with an IFS on sky is a complex process, requiring careful calibrations to ensure that the data can be properly reconstructed into a data cube with good fidelity in both the spatial and spectral directions. Furthermore, observing in the infrared introduces additional challenges associated with telluric effects. These include significant water absorption bands in the operating wavelength band and bright spectrally unresolved sky lines from hydroxyl (OH) airglow. In this section, we discuss calibration and observation strategies we use for WIFIS to obtain good quality data.
\subsection{Calibration}
\begin{figure}
\centering
\includegraphics[width=17cm]{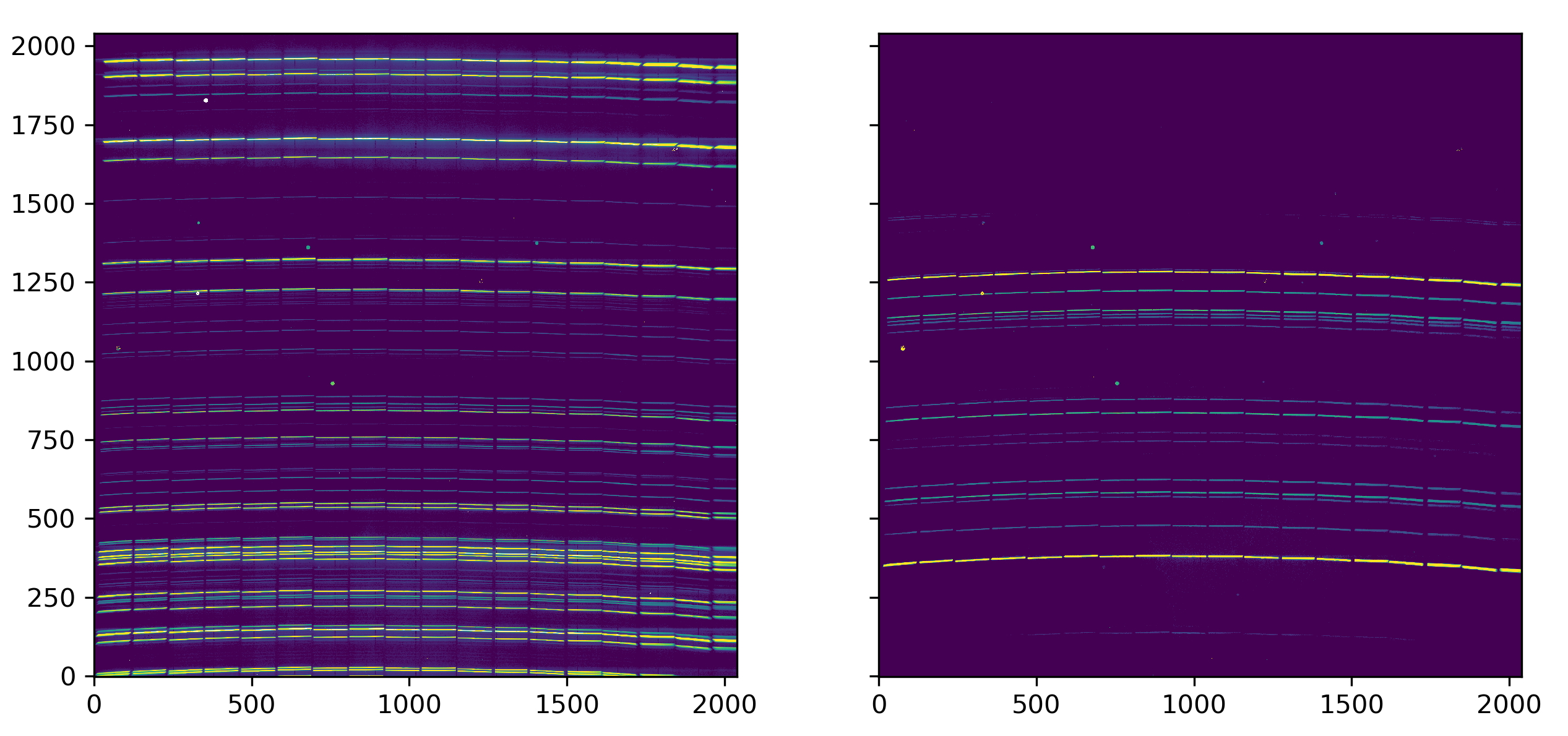}
\caption{Flux images taken with WIFIS illuminated by the ThAr gas discharge lamp in order to obtain wavelength calibration data. The dispersion axis is the y-axis of the images. {\it Left:} Data taken for the $zJ$-band. {\it Right:} Data taken for $H_{short}$ band. A number of interesting features are visible in both images. First, the pseudoslit is curved. This is the ``smile" of the spectrograph. Second, one observes smaller sections in the pseudoslit; these are the individual slices. Each slice has some tilt within the overall slit, requiring each slice to have its own wavelength calibration as a function of spatial position along the slice. Typical RMS values for wavelength calibration residuals are around 0.1 pixel.}
\label{fig:arc}
\end{figure}

The calibration process needs to address both wavelength and spatial calibrations effectively. To accomplish this we have two different light sources in the CU that illuminate the entire WIFIS field. The flatfielding source is designed to produce very uniform illumination across WIFIS's input focal plane in order to flatfield the instrument both spatially and spectrally. The ThAr gas discharge lamp produces relatively uniform illumination of the WIFIS field and is used to determine the wavelength calibration for each slice along its spatial direction. Figure \ref{fig:arc} shows the flux images, slope images generated by fitting up-the-ramp (UTR) exposures, for arcs taken in the $zJ$ and $H_{short}$ bands. Each specific emission line reveals an image of the pseudoslit, which has complex structure, unlike a more traditional long slit. The pseudoslit consists of individual slices arranged in a line, but each slice as a slight tilt with respect to its neighbour. Careful line fitting needs to be done for each slice in order to extract the mapping between wavelength, slice number, and slice position. The arcs are used to find the best internal focus of the instrument. 
\par
There are two different foci within the instrument and we use different methods to obtain the best focus on both. The first focal plane is the WIFIS input focal plane, which is shown at the very top of the instrument in Figure \ref{fig:layout}. This focal plane needs to coincide with the telescope focus. The second focal plane is the spectrograph (internal) focus at the H2RG detector. Given that our CU is designed to be conjugated to the WIFIS input focal plane, we first achieve best internal focus by measuring the full-width-half-max (FWHM) of the pseudoslit over the full wavelength range using arcs while stepping through the detector's focus position. We choose the detector position that produces the best overall focus, i.e., lowest median FWHM for arcs across the entire field. This is done prior to the start of the observing run. The internal focal position remains stable throughout a given run. We discuss how we achieve the proper telescope focus in Section \ref{sec:observing}.
\par
\begin{figure}
\centering
\includegraphics[width=17cm]{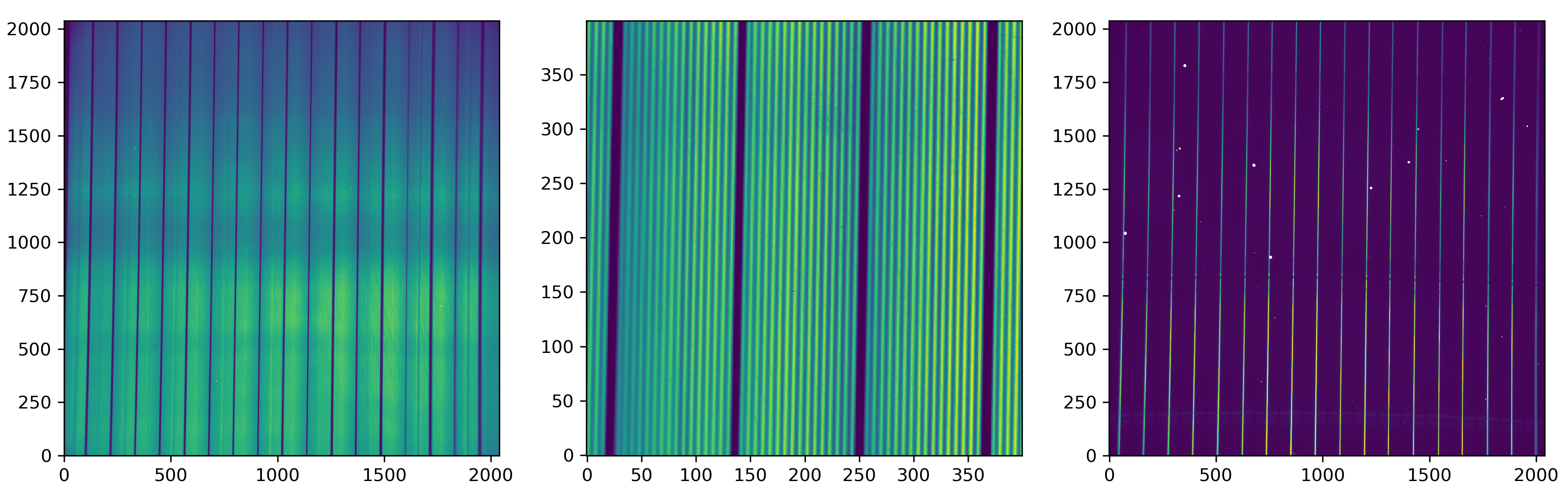}
\caption{WIFIS flux images used for flatfielding and spatial calibrations. The dispersion axis is the $y$-axis of the images. {\it Left:} $zJ$-band flat field image. The slice edges as a function of wavelength are clearly defined by the observed inter-slice gaps. The slice gaps define the edges of individual slices and are used to define the spatial extent of each individual slice. A few broad horizontal bands are observed, which are associated with the variations in grating efficiency as a function of wavelength. {\it Center:} A $400\times400$ pixel section of a flatfield image taken with a Ronchi mask inserted into the output port of the integrating sphere. These data are used to carry out the spatial distortion correction for each slice. {\it Right:} An image derived from an observation of a bright star ($J=3$ mag) scanned across the center of each slice while integrating. This observation provides information to spatially register each slice with one another.}
\label{fig:flat}
\end{figure}

The flatfielding corrects for system throughput variations in position and wavelength due to slice reflectance variations, optical element throughput, grating efficiency, and pixel gain variations, which are substantial in infrared arrays. The flat field image is shown in the left panel of Figure \ref{fig:flat}. One can observe the vertical gaps, which indicate slice boundaries. The flat fields are used to also define slice boundaries as a function of wavelength in the data reduction pipeline (discussed in more detail below). When the object is tracked across the sky, we observe a small motion in the location of the slice boundaries in the flats. The observed flexure is around $2-5$ pixels, which is slightly higher than the original prediction of $<1$ pixel. This is most likely due to flexure within the cryostat that is causing the motion of the WIFIS focal plane array with respect to the rest of the warm optics. The flats are used in the pipeline reduction process to account for the effects flexure in science exposures from a given observation. Finally, we carry out an additional step to determine the spatial mapping along a slice. For each observing run, we insert a Ronchi grating in the output port of the flatfield source, which is then imaged onto the WIFIS focal plane. The Ronchi grating has periodic dark and clear bands that have a very precise width and separation. These bands are oriented across the slices such that every slice is modulated by this pattern. The effect of this is shown in the center panel of Figure \ref{fig:flat}. By tracing these features along the dispersion direction for each slice, we are able to create a 1D spatial distortion map for each slice as a function of wavelength. 


In order to measure the registration between slices, we carry out on-sky calibration observations. A bright star is scanned across the slices along the centre of the field by biasing the tracking rate of the telescope while integrating on source. The result of this is shown in the right panel of Figure \ref{fig:flat}. By tracing the position of the star across the image plane for each slice, the centre position of each slice as a function of wavelength can be determined. We call this measurement the spatial zero point calibration. When combined with the wavelength calibration and one-dimensional (1D) spatial distortion map for each slice, we are able to determine the location of each spaxel at a given wavelength on the detector. These spatial calibration (Ronchi and zero point) measurements are carried out once during an observing run as a routine check for the stability of the mapping. We do not see significant run-to-run variations of the spatial mapping unless we make adjustments to the spectrograph optics, such as realigning the cryostat. 

\subsection{On-Sky Observing}
\label{sec:observing}
Observations are carried out by first acquiring the target in the AGC and centring the object on its field, and then the target is offset into WIFIS's field. This compensates for any residual flexure in the telescope's pointing. If the central position of the requested field does not have a bright object, we use an astrometric technique, which solves for the on-sky position of the AGC, to centre the field on WIFIS. This requires a minimum of three relatively bright stars in the AGC field. Next, a focussing sequence is carried out where the telescope focus is matched to the input focus of WIFIS. This is done by stepping through the telescope focus position while observing a bright star until a minimum number of slices is illuminated. The spectral image is then reconstructed to output a continuum image of the WIFIS field in order to make the final measurement of the size of the point spread function (PSF). Typical measurements reveal a FWHM of around $1.5-2$\arcsec, which is the expected seeing at the Bok telescope site.
\par
After an acceptable telescope focus is achieved, an observing sequence with a certain number of nods is started for an object of interest. A number of key steps occur during an observing sequence: 1) An arc and a flat are taken; 2) The AGC automatically acquires a bright guide star in its offset field and begins guiding. If a sufficiently bright star is not visible, the guide exposure integration time is increased. 3) WIFIS integrates on source for a fixed amount of time (with a maximum observing time of 5 minutes at a time to minimize sky variations). We use UTR integrations for all of our scientific observations. 4) WIFIS nods the telescope to a sky position, usually a few arcminutes away from the source, 5) The AGC acquires another guide star at this new position. 6) WIFIS obtains an off-source measurement of the background with a similar integration time as the object. If the source is a point source, we nod the object within the WIFIS field in order to maximize observing efficiency. We typically follow an ABA sequence for WIFIS observations. 7) WIFIS nods back to the source position and the AGC acquires the same guide star and moves it to its original position and begins guiding as WIFIS continues integrating on source. 8) This sequence continues until the requested number of nods is completed.
\par
\begin{figure}
\centering
\includegraphics[width=17cm]{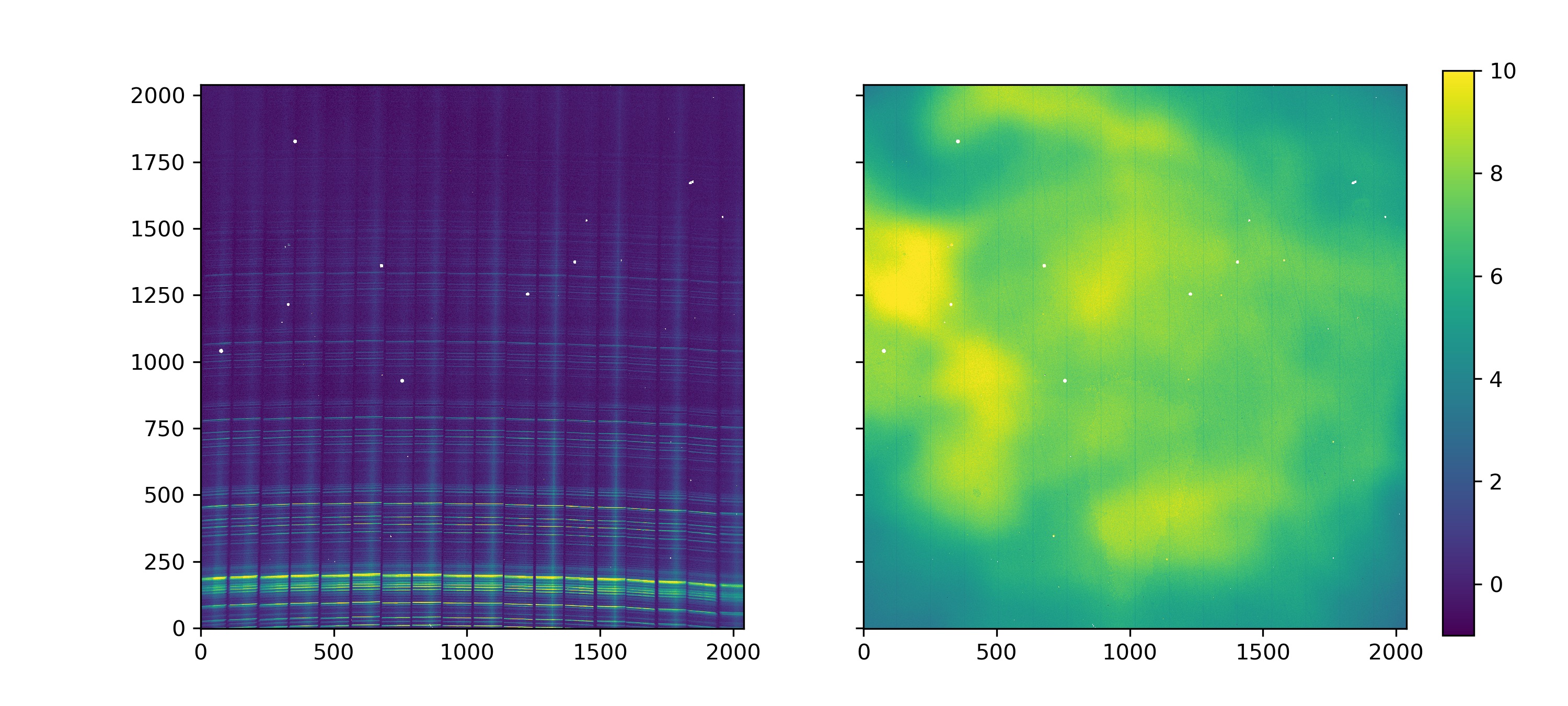}
\caption{{\it Left:} Flux image of Messier 87, a giant elliptical galaxy, generated from a 5-minute exposure consisting of 200 UTR reads. The dispersion direction is along the $y$-axis. The central slice in WIFIS's field-of-view is located on the very left. The flux scale bar in units of e$^{-}$ s$^{-1}$ is given in the far right. The thermal background on this image has been subtracted. The sky emission lines are clearly visible and follow the shape of the pseudoslit. The continuum emission of the galaxy is seen as a vertical line in multiple slices. The white dots show regions of inoperable pixels. {\it Right:} Thermal background flux image generated from several 5-minute exposures averaged together. The flux scale bar in units of e$^{-}$ s$^{-1}$ is given in the far right. This background is undispersed and shows non-uniform illumination most likely due to variations in long wavelength cut-off of the detector.}
\label{fig:m87thermal}
\end{figure}

Sky subtraction cannot be effectively done without additional sky pointings because of the non-uniform thermal background and slight differences in the responsivity of the slices that cannot be easily flatfielded. Additional arcs and flats are taken every 45 minutes of observations for a given object. This is required to track the effects of flexure in the focal plane of the instrument. For faint extended objects, we take 5-minute long individual exposures on source and observe them for approximately two hours, including sky observations. A typical observation of a faint extended source, Messier 87 in this case, is shown in the left panel of Figure \ref{fig:m87thermal}. The thermal background that has been subtracted from that image is shown in the right panel of the figure. Telluric observations of A0V stars are typically done before and after each observation. These stars are chosen to be located at the average airmass of the source during its observing sequence. In addition to removing telluric features, these stars are used to flux calibrate the data.

\section{DATA PIPELINE}
\label{sec:pipeline}
\begin{figure}
\centering
\includegraphics[width=15cm]{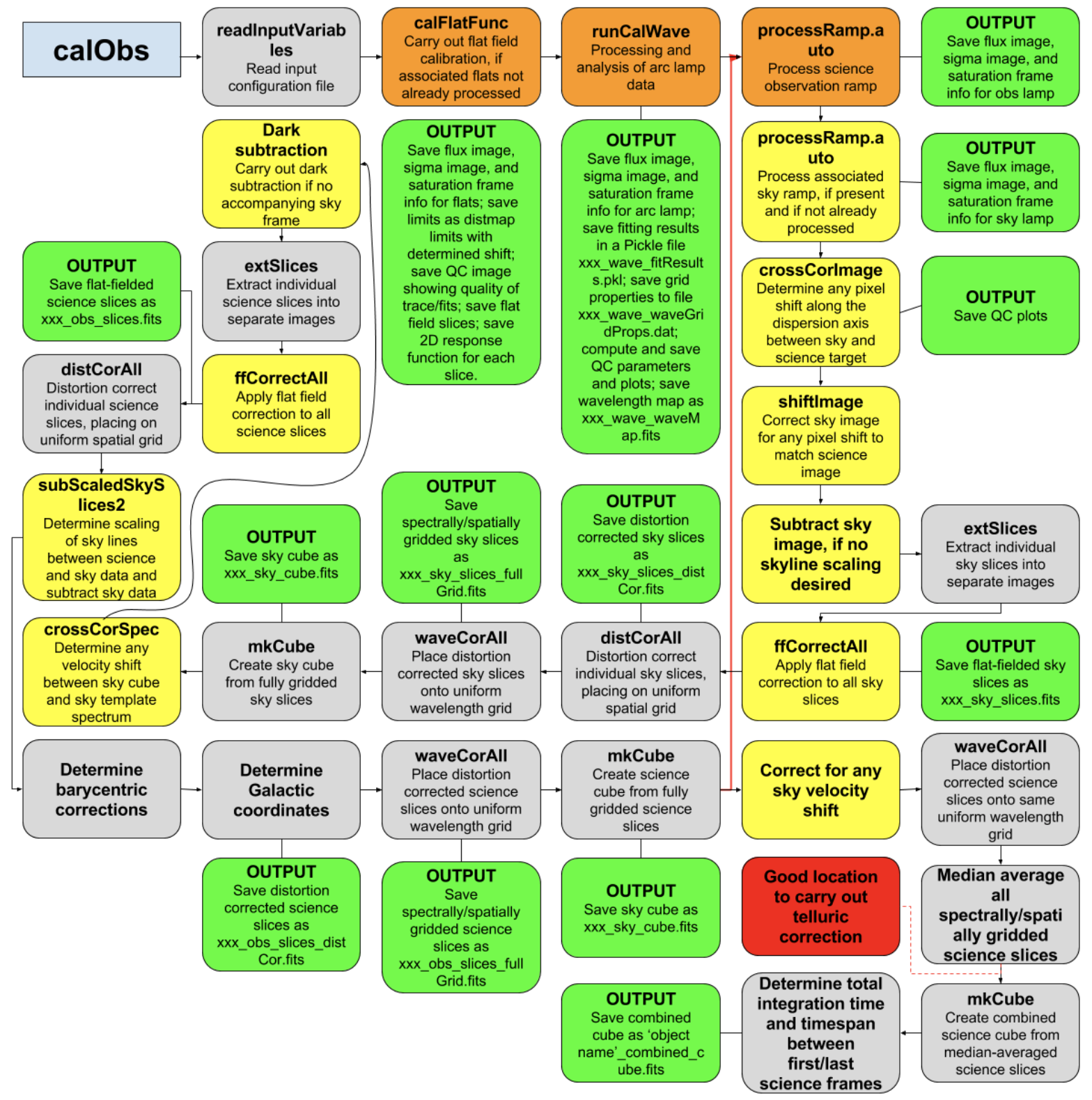}
\caption{Flowchart of the data reduction process for a science observation. A single script called \emph{calObs} with the appropriate input parameters is invoked and with minimal user interaction a spectral cube is generated at the end of the process. Automatic telluric corrections are currently not implemented in the pipeline, but will be in a future date.}
\label{fig:pipeline}
\end{figure}

\begin{figure}
\centering
\includegraphics[width=11cm]{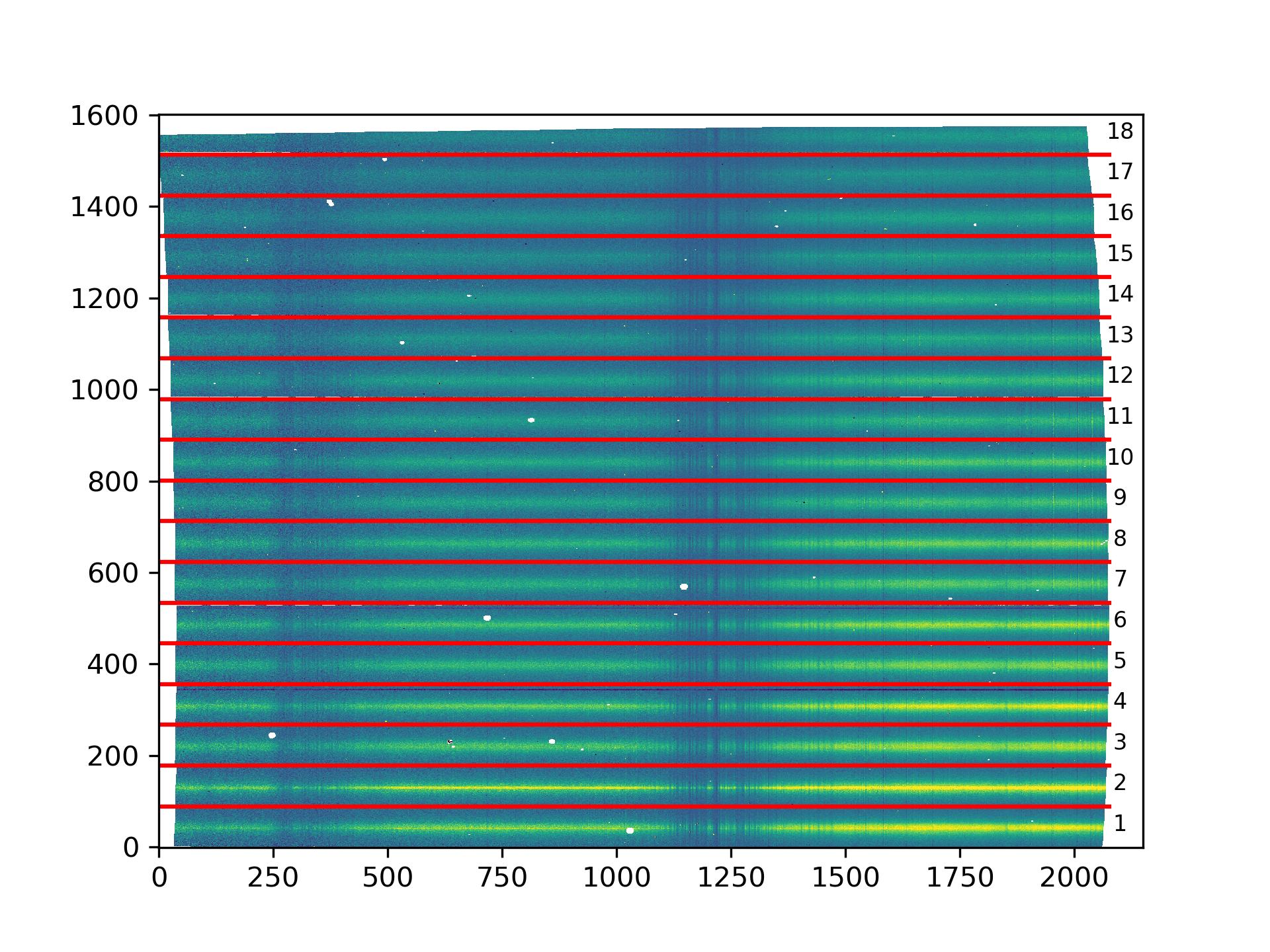}\raisebox{2.5cm}{\includegraphics[width=6cm]{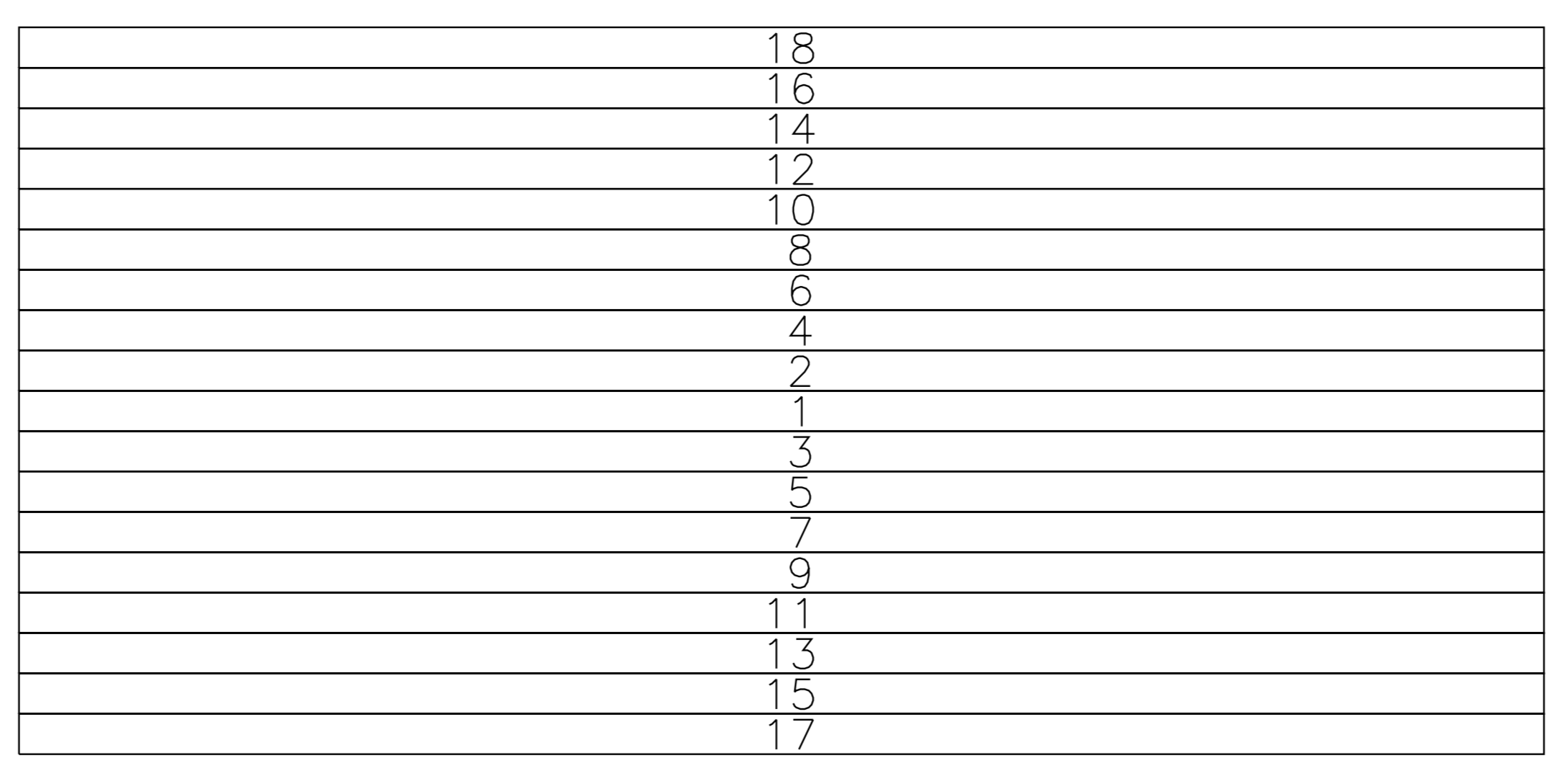}}
\caption{Pipeline processed flux image of individual slices for an 5-minute observation of M87 is shown in the left panel. The individual slices are numbered. The central slice is the first slice, and the slices alternate as one moves further from the centre of WIFIS's field. A mapping of the slices is shown in the right panel, which is discussed in greater detail in an earlier paper\cite{sivanandam2012}. The data have been flatfielded, background subtracted and placed on a uniform wavelength and spatial grid after wavelength and spatial calibrations. The dispersion axis is along the x-axis with the shortest wavelength on the left. The final step is to spatially align these slice data with each other and generate a spectral cube. The continuum emission of the galaxy is clearly visible in all slices as well as the strong telluric absorption features. }
\label{fig:m87pipeline}
\end{figure}

The WIFIS data reduction pipeline (PyPline) was written from the ground up in \emph{Python} and takes advantage of OpenCL and multiprocessing to significantly speed up calculations. The software is open-source and is available at the following link: \url{https://github.com/WIFIS-Team/pipeline}. The current implementation of the pipeline is fully featured and can take raw frames and produce spatially and wavelength calibrated data cubes with relative ease, provided good calibrations are available. Key features include reference pixel corrections and non-linearity corrections to H2RG data, ramp fitting, flatfielding, automatic wavelength solution fitting, and spatial reconstruction. Additional features include automatic sky and background subtraction, flexure compensation, and multi-observation averaging. The pipeline is able to compensate for subpixel shifts in the wavelength solution of each individual exposure by tracing minor shifts in the OH lines, as well as scaling the brightness of these sky lines to best subtract them from the science data. Flexure compensation is done in post processing by tracking the movement of slice edges in calibration flats. Telluric correction is currently not implemented in the pipeline and is carried out manually. The flowchart of the pipeline's functionality is shown in Figure \ref{fig:pipeline} where \emph{calObs} is the script that is called to reduce science data. An example of pipeline processed flux image of M87 that has spatial and wavelength calibration applied is shown in Figure \ref{fig:m87pipeline}. This figure shows the data that is used to construct the final sky subtracted datacube. The arrangement of slices on sky is shown in the right panel. The pipeline is highly optimized for speed and can reduce a two hour observation ($\sim30$ GB of raw data) in 20 minutes. A more detailed paper on the methods used in the pipeline and its overall performance will be presented at a later date.

\section{ON-SKY PERFORMANCE}
\label{sec:performance}
\begin{figure}
\centering
\includegraphics[width=8.4cm]{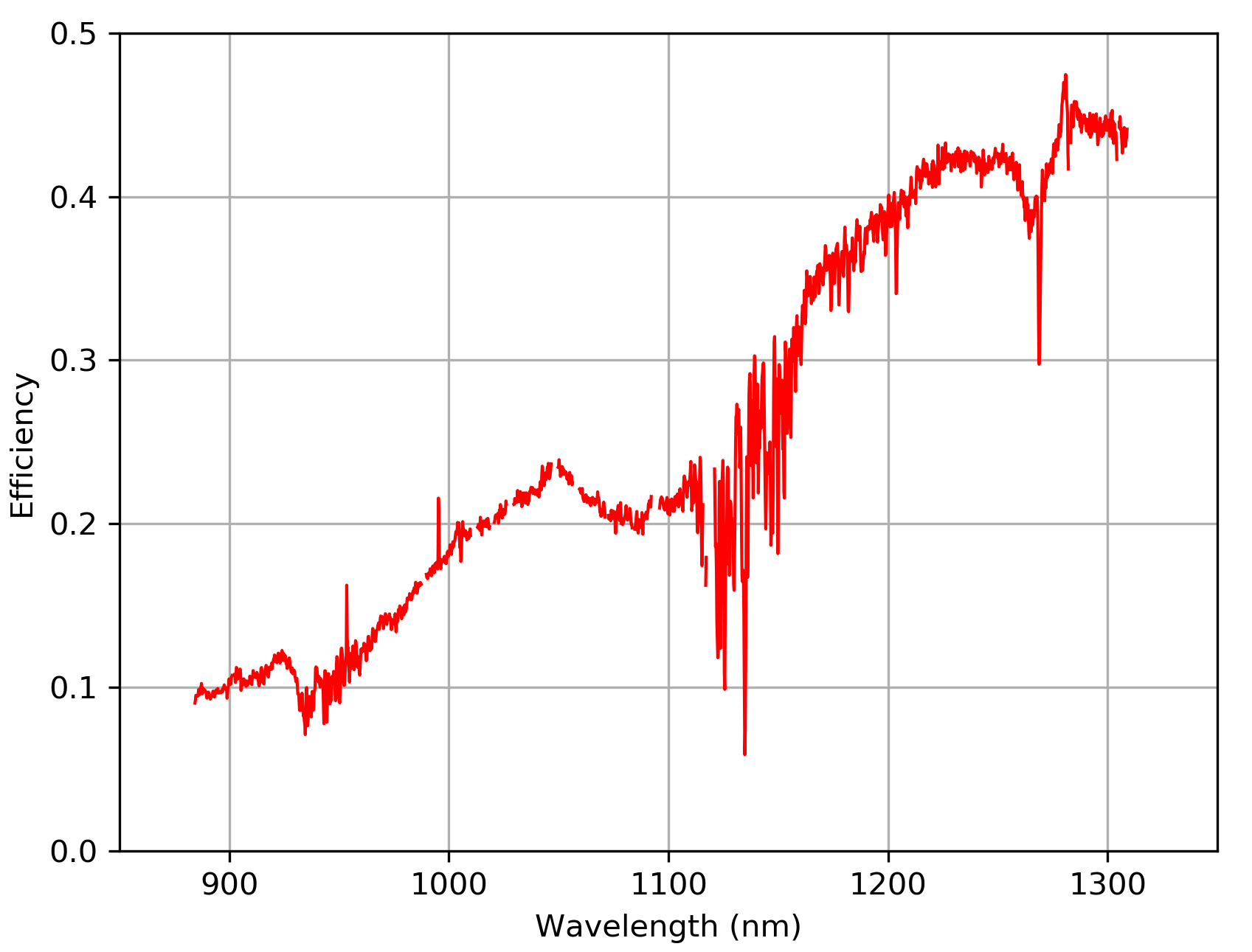}\includegraphics[width=8.6cm]{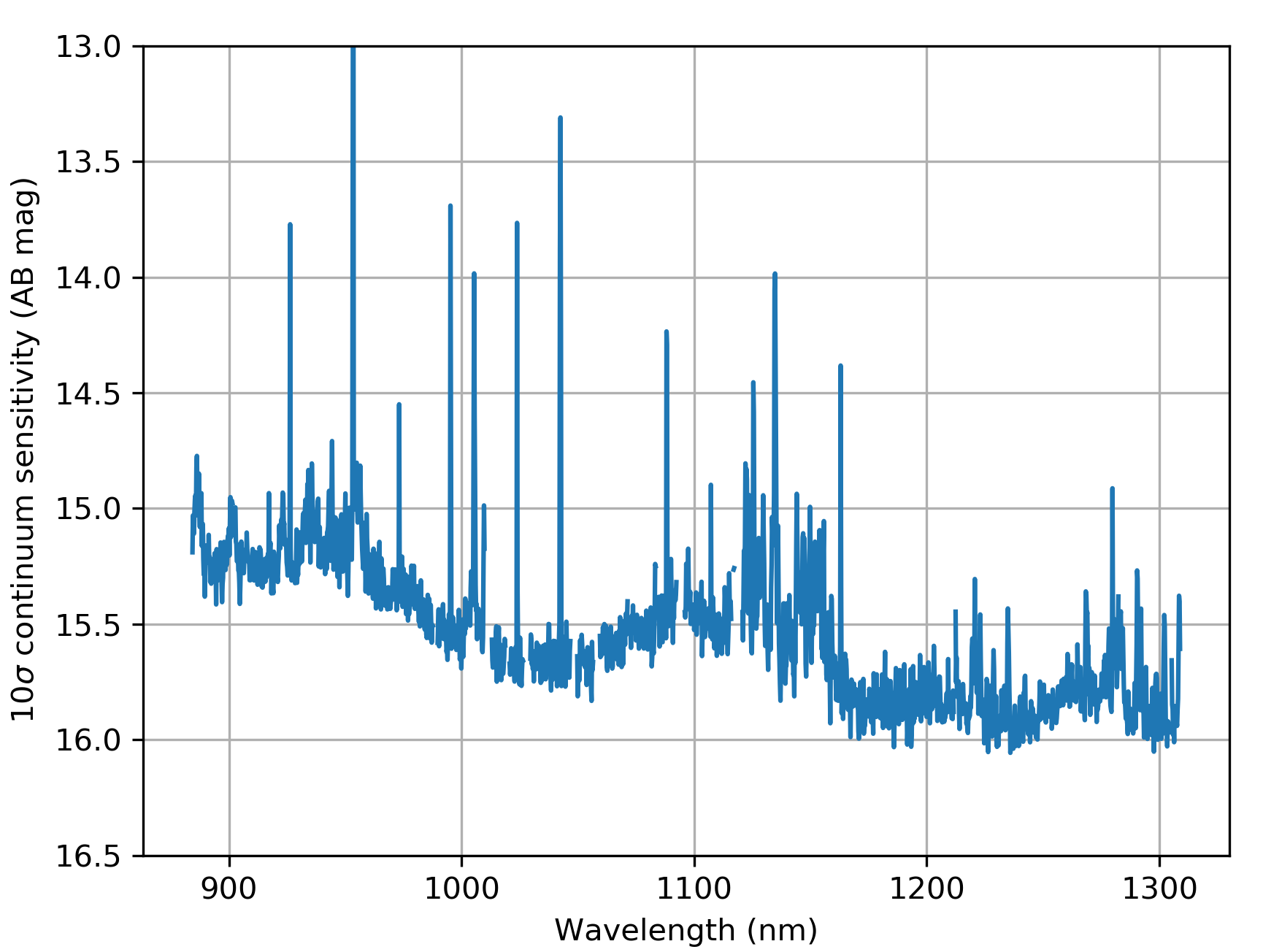}
\caption{{\it Left:} Overall end-to-end $zJ$-band efficiency measurement for WIFIS under typical conditions (seeing and precipitable water vapour) at the Kitt Peak 2.3-meter telescope derived from telluric star observations. This measurement includes the throughput of the atmosphere and the telescope. The shape of the curve is largely dominated by the diffraction efficiency of the grating, which is blazed at 1370nm. Telluric absorption bands are observable at 950 and 1150nm. {\it Right:} 10$\sigma$ point source continuum sensitivity for a 1 hour on-source integration. This sensitivity estimation includes an additional 30 minute sky observation.}
\label{fig:performance}
\end{figure}

The instrument saw first light in May 2017 and its performance has been steadily optimized over the following runs. We have taken numerous datasets to characterize WIFIS's on-sky performance. Through telluric observations of $J=9$ mag star, we were able to measure the overall throughput of the instrument, which is shown in the left panel of Figure \ref{fig:performance}. This includes all of the losses in the optical train, including the atmosphere and the telescope. The overall throughput of WIFIS ranges from $10-40$\% across the $zJ$-band. This is largely dominated by the grating efficiency. The grating was chosen to be useable in both the $zJ$ and $H_{short}$ bands; consequently, it was blazed at 1.37$\mu$m. Given that WIFIS is not very sensitive in the $H_{short}$-band due to its significant thermal background, it will be mostly used in the $zJ$-band. A grating with a different blaze angle could provide a modest increase in the throughput and sensitivity at the bluer wavelengths. The point source continuum sensitivity for a one-hour on-source integration is presented in the right panel of Figure \ref{fig:performance}. This was derived by measuring the noise in source-free regions in the pipeline-reduced, sky-subtracted cubes. This sensitivity estimate was derived using a $\sim 3$ minute integration of $J = 9$ mag A0V telluric star, which was observed using a single ABA sequence. The measured sensitivity was then rescaled to a one hour on-source exposure. WIFIS reaches a 10$\sigma$ sensitivity of 16 AB mag with approximately a 1 mag variation between the blue and red end of the $zJ$-band.  This variation can be explained by the increased throughput in the red end. The sensitivity measurement is approximately $1.5-2$ mag worse than our original predictions, which is mainly because the measured thermal background of $2-10$ e$^{-}$ s$^{-1}$ pix$^{-1}$ is about an order of magnitude higher than our original specification of $0.5$ e$^{-}$ s$^{-1}$ pix$^{-1}$. We discuss future thermal background mitigation strategies later.

\begin{figure}
\centering
\includegraphics[width=5cm]{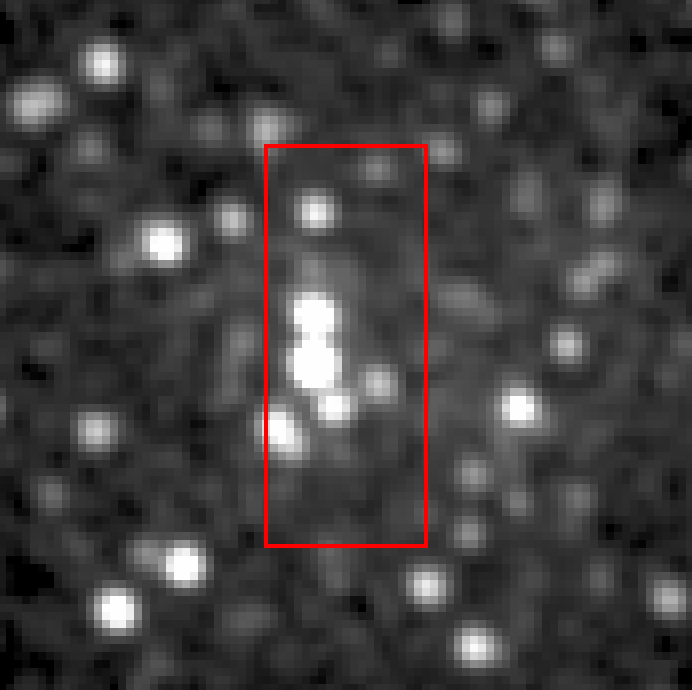}\hspace{1cm} \raisebox{0.95cm}{ \includegraphics[height=3cm]{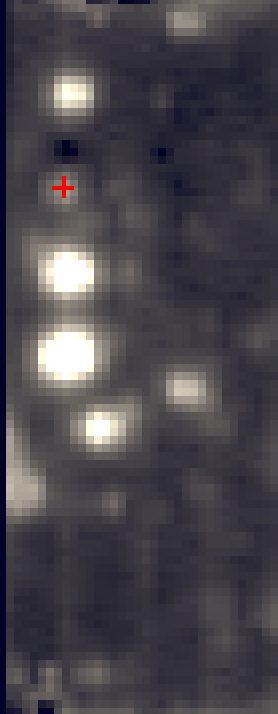}}
\caption{{\it Left:}  2MASS $2\times2$\arcmin $J$-band image of NGC6426, an old Milky Way globular cluster. The red rectangle shows the WIFIS observing field. {\it Right:} A continuum flux image in the $zJ$-band derived from WIFIS data for region outlined in the 2MASS image. The integration was 1.5 hours on-source. The cross marks the detection of a $J_{AB}=16$ mag star in the cluster. This observation shows the imaging capability of WIFIS.}
\label{fig:ngc6426}
\end{figure}

\begin{figure}
\centering
\includegraphics[width=7cm]{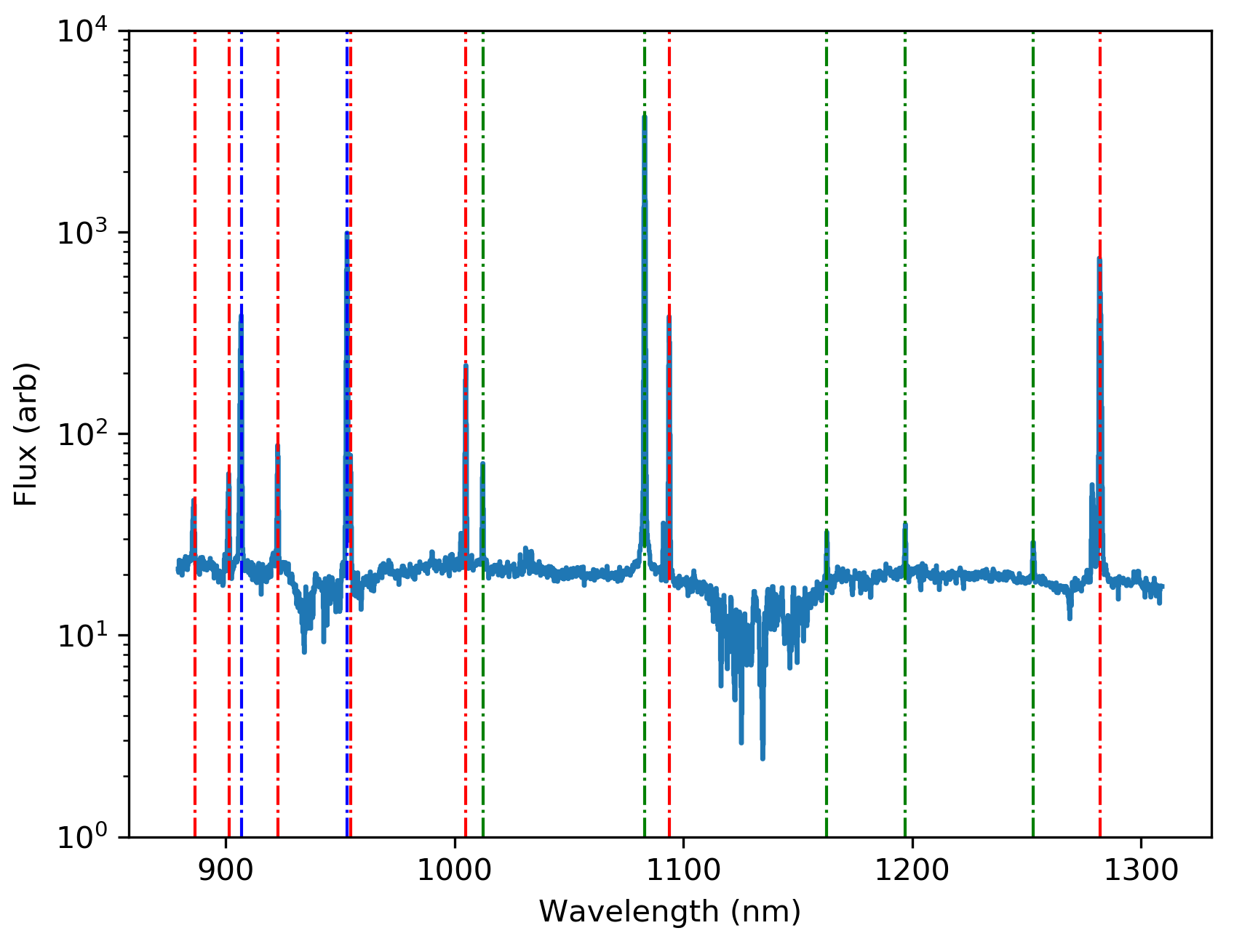}\hspace{1cm}\raisebox{0.6cm}{\includegraphics[height=4.5cm]{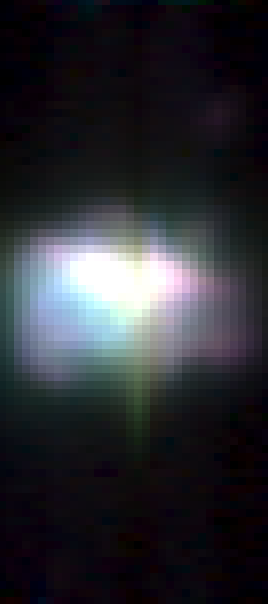}}
\caption{{\it Left:} Spectrum of NGC6210, a planetary nebula, extracted from a 2\arcsec aperture centred on its white dwarf. The continuum emission and the nebular lines are clearly visible. The red vertical lines indicate emission from the Hydrogen Paschen series, while the green are Helium I lines, and the blue are forbidden sulphur [SIII] features. The absorption features observed at 950 and 1150 nm are telluric features. {\it Right:} Image reconstruction of NGC6210. The red, green, and blue channels trace the Pa$\beta$, He I 1083nm, and [SIII] 953.2nm lines, respectively. Variations in line emission across the PNe is clearly visible in these observations, as evidenced by differences in colour across the PNe.}
\label{fig:ngc6210}
\end{figure}

We present a few example observations of astrophysical objects, a globular cluster, a planetary nebula, and a nearby galaxy, below. In Figure \ref{fig:ngc6426}, we show an observation of the central region of a nearby globular cluster, NGC 6426. This observation demonstrates WIFIS's capability to perform imaging observations. The right panel of the image is a continuum image derived from collapsing the spectral cube in the wavelength direction. Given that we integrated 1.5 hours on-source, this observation demonstrates that we can guide effectively with our AGC and maintain good image quality with WIFIS. In Figure \ref{fig:ngc6210}, we show a spectrum and an image reconstruction of a planetary nebula (PNe), NGC 6210. This PNe is known to exhibit highly asymmetric structure and has a physical size that is well-matched to WIFIS. The spectrum is extracted from a region centred on the white dwarf within the PNe. Along with several strong emission lines, the continuum emission from the white dwarf is also visible. Emission lines from several different elements, hydrogen (H), helium (He), and sulphur (S), are detected. The Paschen ladder is visible, along with a few He I lines, as well as [SIII] lines. The three colour image of the PNe highlights emission from [SIII], He I, and Pa$\beta$, which are depicted by blue, green, and red colours, respectively. The ionization structure of each line varies across the PNe, likely indicating abundance variations.

\begin{figure}
\centering
\includegraphics[width=8.5cm]{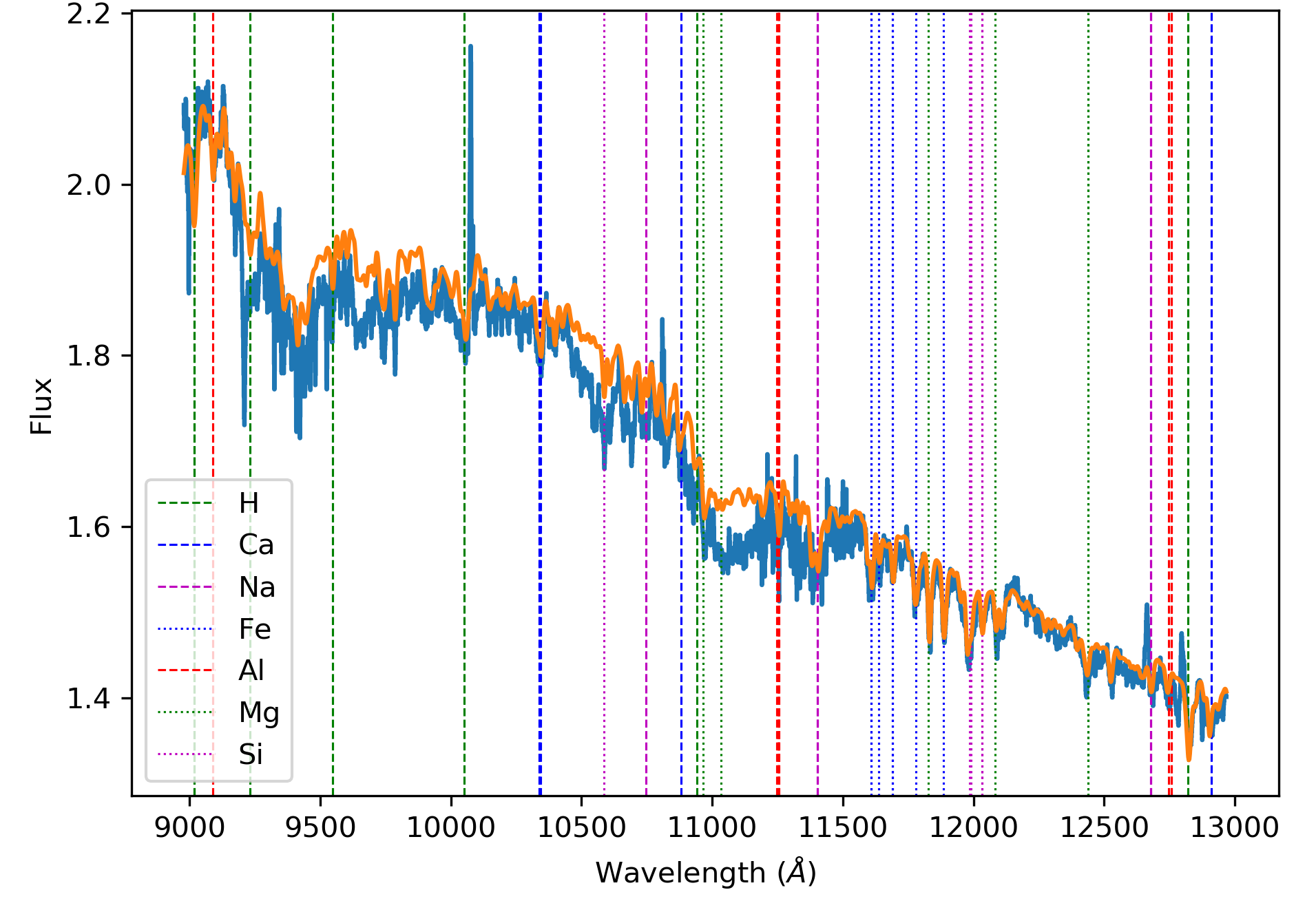}\hspace{1cm}\raisebox{0.8cm}{\includegraphics[height=4.5cm]{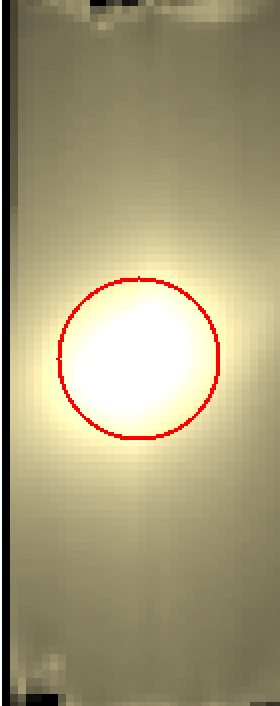}}
\caption{{\it Left:} Fully reduced WIFIS spectrum of M85 integrated within the central 10\arcsec of the galaxy. The total on-source integration time was 60 minutes. The blue curve is the telluric corrected spectrum of the galaxy, whereas the orange curve is a stellar population synthesis (SPS) model at the age of 3.0 Gyr\cite{conroy2012}. The flux of the SPS model is rescaled to match the data at 1.2 $\mu$m. After rescaling, the model matches the data well. Many of the key lines for IMF studies are clearly detected in the WIFIS spectrum. The forest of lines observed in the 1.15-1.25 $\mu$m range are especially useful for measuring stellar kinematics. {\it Right:} Continuum image of M85 in the $zJ$-band reconstructed from WIFIS data. The spectral extraction region is the red circle.}
\label{fig:m85}
\end{figure}

Given that we are carrying out a large survey of nearby galaxies to study their stellar populations, we present preliminary results of Messier 85 in Figure \ref{fig:m85}. The figure shows a telluric corrected spectrum of the galaxy within its central 10\arcsec generated from one hour of data on-source and a reconstructed continuum image with the extraction region highlighted. Overplotted on the spectrum is a model spectrum that matches the age and metallicity of the galaxy generated using a spectral synthesis model\cite{conroy2012}. The model spectrum has only been scaled by a multiplicative factor to match the spectrum at 1.2 $\mu$m. The model agreement is very good. Several detected metal and hydrogen features are highlighted in the spectrum. The final goal of our scientific program is to IMF sensitive features and stellar kinematics of the galaxies in our sample to determine the giant-to-dwarf ratio and mass-to-light ratio of these systems in order to constrain and possibly detect variations in their IMFs.

A few outstanding performance issues remain with WIFIS. The most pressing is the thermal background, which is $4-20$ times higher than our original specification in the $zJ$-band. The large variation is due to changes in ambient temperatures. The background is expected to double with every 10C increase in temperature, which was observed during the different runs with varying ambient temperatures. Furthermore, the background is not uniform across the camera field, which we surmise to be variations in the long wavelength cut-off of the detector. This could result from differences in the Urbach tail across the detector. The reason for the elevated background was because our custom interference filter was not designed to be sufficiently aggressive in reducing thermal background entering the cryostat at high incidence angles. In order to determine the feasibility of reducing the background further, we carried out a test by inserting our spare blocking filter in series with the existing one. Given that these filters have very high in-band throughput of $\sim 99$\%, there was no appreciable loss in light. The thermal background, however, was reduced by a factor of $5-8$, which largely meets our original specification. Given an additional filter adds significant optical path length, we were unable to use the instrument in this configuration. Instead, we are in the process of acquiring a new set of filters that can provide improved blocking performance.  A better filter, with improved thermal blocking at higher angles of incidence, would yield significant improvements in the overall sensitivity and could allow us to reach our original design performance.
\par
\begin{figure}
\centering
\includegraphics[width=10cm]{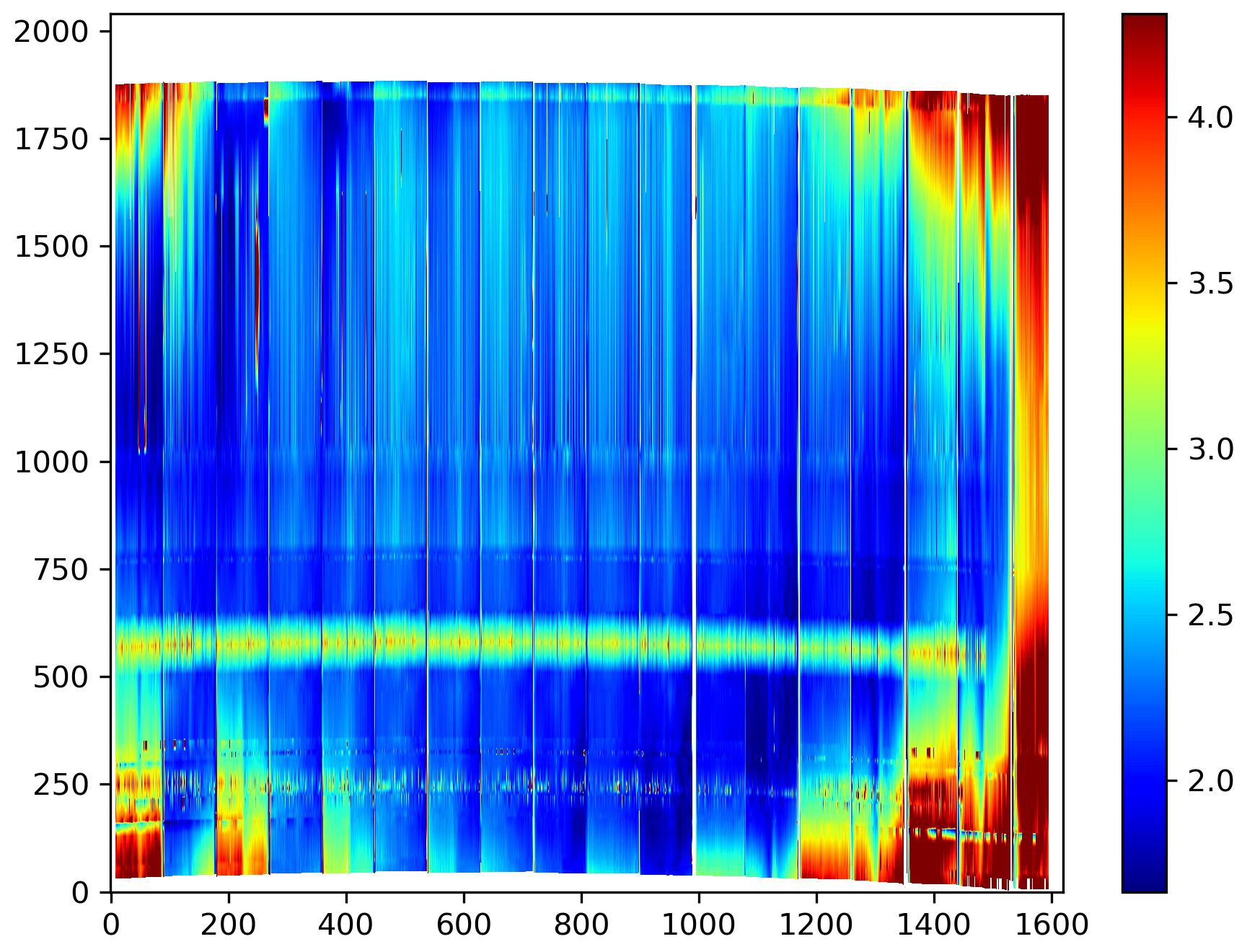}
\caption{FWHM map derived from arc images of WIFIS's slices in the $zJ$-band. The map is generated by interpolating the FWHM of the detected ThAr lines across the full field of the spectrograph camera. The colour scale is in pixels where the median FWHM in this image is 2.3 pixels. The dispersion direction is along the $y$-axis while the spatial direction is along the $x$-axis. The increased FWHM along $y=600$ is an artefact due to a saturated line. The degradation of the performance along the edges of the spectrograph camera field is clearly visible with the edge slices on the far right suffering the most.}
\label{fig:specperf}
\end{figure}

The second issue is the degradation of image quality at the edges of the spectrograph camera field. This is shown in Figure \ref{fig:specperf}, which is a map of the FWHM measured from arc images. Given that the spectral and spatial foci are designed to be coincident on the instrument's focal plane array, a degradation in the FWHM in the spectral direction has the same effect in the spatial direction. The central slice, which is the left most slice in Figure \ref{fig:specperf}, has reduced image quality and spectral resolution at the shortest and the longest wavelengths while the very edge slices, located on the right side of this figure, have poor image quality at all wavelengths. This is likely due to the misalignment of the last optical element, the field flattener lens, or a slight tilt in the detector with respect to the overall optical system. More work needs to be done to fully characterize and correct the problem. 

\section{CONCLUSIONS}
\label{sec:conclusions}
We present the overall construction and performance of a novel wide integral field infrared spectrograph. WIFIS offers the widest field possible on a seeing-limited infrared integral field spectrograph with an etendue that is unrivalled by any existing infrared IFSes. With a field-of-view comparable to existing visible-light IFSes, WIFIS can be an excellent follow-up instrument for on-going optical surveys of very extended objects such as supernova remnants and nearby galaxies. WIFIS largely meets its original performance goals of carrying out $R\sim3,000$ imaging spectroscopy in the $zJ$ and $H_{short}$ bands. However, its overall sensitivity can be further improved by reducing the thermal background, which currently exceeds our design goals by up to an order of magnitude. 

\acknowledgments 
 
We acknowledge our funders, the Canada Foundation for Innovation, the Ontario Research Fund, the Dunlap Institute for Astronomy and Astrophysics, and the Korea Astronomy and Space Science Institute. We thank Ken Blanchard for his important contributions during the design and manufacture of WIFIS components. We are very appreciative of the Steward Observatory staff, Joseph Horscheidt, Melanie Waidanz, William Wood, and Kenneth Don for their contributions in getting WIFIS operational at the Kitt Peak Bok telescope. We also thank Margaret Ikape who developed the telluric correction software for removing telluric absorption features from WIFIS M85 data.

\bibliography{report} 
\bibliographystyle{spiebib} 

\end{document}